\documentstyle[12pt]{article}
\newcommand{\be}{\begin{equation}}
\newcommand{\ee}{\end{equation}}
\newcommand{\ba}{\begin{eqnarray}}
\newcommand{\ea}{\end{eqnarray}}
\newcommand{\baa}{\begin{eqnarray*}}
\newcommand{\eaa}{\end{eqnarray*}}
\newcommand{\bb}{\begin{thebibliography}}
\newcommand{\eb}{\end{thebibliography}}
\newcommand{\ci}[1]{\cite{#1}}
\newcommand{\bi}[1]{\bibitem{#1}}
\newcommand{\lab}[1]{\label{#1}}

\begin{document}

\begin{center}
\phantom{.}
\begin{flushright}
JINR Preprint E2-94-348\\
Dubna,1994
\end{flushright}

\vspace {3cm}
{ SPIN EFFECTS IN HIGH ENERGY PHOTON-HADRON SCATTERING IN QCD}\\
S.V.Goloskokov,
\begin{footnote}
{ E-mail:goloskkv@thsun1.jinr.dubna.su}
\end{footnote}
 O.A.Listopadov
\begin{footnote}
{ Scorina University, Gomel, Belarus}
\end{footnote}
\\
Laboratory of Theoretical Physics,\\
Joint Institute for Nuclear Research, Dubna\\
\end{center}
\begin{abstract}
 The spin effects at high energies and momenta transferred  $|t|>1 GeV^2$ are
analysed for quark-photon scattering. The energy independence of the ratio
of spin-flip and spin-non-flip amplitudes in the same $\alpha_s$ order of
QCD is obtained. It is shown that the contribution of the spin-dependent
quark-pomeron vertex to the spin-flip amplitude is intensified by
off-mass-shell effects in the quark loop. As a result, the magnitude of this
amplitude can reach 20-30\% from a spin-non-flip one. The dependence of
the cross-section on the form-factor and $\alpha_s^3$ spin-non-flip
contributions are observed.
\end{abstract}

\newpage
\phantom{.}
\vspace {1.0cm}

At present, the study of spin effects has attracted considerable interest
due to the development of polarized programmes at future accelerators
\ci{bunc}.
Most part of the spin experimental data at high energies is now obtained at
fixed momenta transferred. The $t$-channel exchange with vacuum quantum
numbers (pomeron) gives the main contribution to this region.
The vacuum $t$-channel amplitude is usually associated in QCD with the
two-gluon exchange \ci{low}. The spinless pomeron was analysed
in \ci{la-na,don-la} on the basis of a nonperturbative QCD model.
 A similar model was used
to investigate the spin effects in the pomeron exchange.
It has been shown that different contributions like a gluon \ci{gol1}
and quark loops \ci{gol2,gol4} may lead to the spin-flip amplitude
growing as s in the limit $s \to \infty$. As a result, the spin-flip
amplitudes are suppressed  logarithmically with respect to the
spin-non-flip amplitude of the same order in $\alpha_s$:

\be
\frac{\vert\; T_{f}\;\vert }{\vert \;T_{nf}\;\vert} \simeq
\frac{m \sqrt{\vert t \vert}}{a(m,t)\,\ln{s}}.   \lab{1}
\ee
Here and in what follows $m =0.33 GeV$ is the constituent quark mass and $a$
is a function linearly dependent on  $\vert t\vert$  at large $\vert t\vert $.
The arise of $\ln{s}$ in (1) stems from the fact that the
spin-non-flip amplitude has an additional logarithmical factor.
Just this amplitude was calculated in different papers (see \ci{wu} e.g.).
This result confirms the absence of spin-flip amplitudes in the
leading log approximation \ci{lip}.

 In papers \ci{gol1,gol2,gol4}, the high-energy $qq$ scattering was studied
in the soft momentum transfer region $|t|<1GeV^2$. In  \ci{gol5,go-se},
spin effects were investigated in the semi-hard region
$s \to \infty,\; |t|>1GeV^2$ where the perturbative theory can be used.
 Since the spin-flip amplitude growing as $s$ is absent in the
born two-gluon diagrams Fig. 1a more
complicated ladder diagrams were considered.

 It has been shown that the main contribution to the spin-flip amplitude
comes from the planar diagrams of the form, Fig. 1b, 1c.
Moreover, there are $s \rightarrow u$ crossing diagrams with  crossed
gluon lines. In the semi-hard region we have $u \simeq -s$ and
the real parts are compensated in the sum of diagrams. So we can calculate
only the imaginary parts of diagrams in the case of the pomeron exchange.
 In \ci{gol5,go-se} the factorization spin-flip amplitude $T_f$ into the
spin-dependent large-distance part and the high-energy spinless pomeron
were shown. This permits one to  define the quark-pomeron vertex which
is appropriate for
the  spin-dependent low-energy subgraphs (the upper parts of the graphs
Fig. 1b, 1c) and for investigating the results of summation of the pomeron
ladder graphs in higher orders of QCD. As a result of this summation,
it has been found that the ratio  $T_{flip}\Big/T_{non-flip}$ is
energy-independent.
The obtained total spin flip amplitude \ci{gol5,go-se}  is about 2 per cent
 of the spin-non-flip one. Note that the smallness of the resulting
spin-flip amplitude is caused by the compensation of different matrix
structure contributions, that are not very small by themselves.
Compensation like that is possible only for the quarks on the mass shell.
Really, the model investigations
of the $\gamma q$ elastic scattering at high energies \ci{gol5,gol4} show that
the off-mass-shell effects in the quark loop increase the spin-flip amplitude
essentially.

The purpose of this paper is to investigate the role of
the off-mass-shell effects of the wave functions in the semi-hard region.
The role of these effects is studied by using  the  elastic $\gamma q$-
scattering as an example. The contributions from the hadron wavefunction are
very similar to the quark-loop integral in this reaction and all calculations
can be performed up to the end without taking any additional information
about the hadron spin structure into account. Moreover, the investigation of
$\gamma q$-scattering is of major methodical importance since it
 is the simplest subprocess in photon-hadron reactions.
For example, the reactions of diffractive photoproduction of vector mesons
 differ from the above only by the replacement of the wavefunction of the
outgoing photon by the meson wavefunction.

  Let us investigate the elastic $\gamma q$-scattering
  $$
\gamma\; (p_1)\;+\;q\;(p_2)\;=\;\gamma\; (p_3)\;+\;q\;(p_4).
$$
In what follows, we shall use the symmetric coordinate system in which
  the sum of quark momenta before and after scattering is directed
along the $z$-axis:
\be
q=\frac{p_1+p_3}{2}=(p_0,0,0,p_z),\;\;\;\;\;
p'=\frac{p_2+p_4}{2}=(p_0,0,0,-p_z),
\ee
and the momenta transferred $\Delta$, along the $x$-axis \ci{gol1}:
\be
r=\frac{p_1-p_3}{2}=\frac{p_2-p_4}{2}=(0,-\Delta/2,0,0).
\ee

 In the present paper only the planar graphs (which contain an interaction
of the gluons from the pomeron
 with one  quark in the loop) will be considered
since just they determine effects under study.
The main contribution to the elastic $\gamma q$-scattering amplitude
in the semi-hard region in the $\alpha_s^2$ order comes from the imaginary
part of the diagram drawn in Fig. 2a.
In calculations, we shall use the Feynman gauge because only the $g_{\mu,\nu}$
terms in the $t$-channel gluon propagators contribute to the leading $\sim s$
terms of the scattering amplitudes, and in the $\alpha_s^3$ order we do not
have ghost contributions.

Let us calculate the matrix element of
the amplitude, Fig. 2a, with  spin-flip and spin-non-flip of the photon.
As has been shown in \ci{gol1}, the main contribution to the amplitude
is determined by the following spin-non-flip matrix element structure
in the lower quark line
\be
\bar u^+(p'+r) \gamma^\mu( \hat p'+ \hat l+m)
 \gamma^\nu u^+(p'-r) \simeq 4 p'^\mu  p'^\nu.    \lab{down}
\ee
Then, the matrix element of the amplitude can be written as follows:
\be
M=N^{\mu,\nu} 4 p'_\mu  p'_\nu ,
\ee
 where $ N_i^{ \mu \nu}$  is a matrix structure of the upper graph's part.

Now let us calculate the imaginary parts of the spin-non-flip matrix elements
in the down quark line $\langle T_{Born}^i(s,t) \rangle$  of
diagram Fig. 2a. They have the form
\ba
{\rm Im}\langle T_{Born}^i(s,t) \rangle=cof \int d^4p d^4l
\delta[(p-q)^2-m^2]\delta[(p-l)^2-m^2]\nonumber\\    \nonumber\\
\delta[(p'+l)^2-m^2]
\left[ M_{Born}^i\right] \; \prod (G(p\pm r))
\prod (F(l\pm r)),
\ea
where $cof$ is a numerical factor, $G(p\pm r)$ is the quark propagator
functions from the upper part of the graph,
$F(l\pm r)$ are the gluon propagators  from the lower part of the graph,
$M_{Born}^i$ include the corresponding matrix elements of the diagram's
numerator.
It is useful to perform calculations in the light-cone variables
\be
p=(zq_+,p_-,p_\perp),\;k=(xq_+,k_-,k_\perp),\; \;l=(yq_+,l_-,l_\perp),\;
q_\pm =q_0 \pm q_z\;.   \lab{lc}
\ee
After integration with
the $\delta$-functions, we obtain the following representation:
\ba
{\rm Im}\langle T_{Born}^i(s,t) \rangle=cof\frac{1}{s}
\int_{s_0/s}^{1}\frac{dz}{z(1-z)} \;\int\;d^2p_\perp  d^2l_\perp \nonumber\\
\nonumber\\
\left[\;M_{Born}^i
\;\prod (G(p\pm r)) \prod (F(l\pm r))\right]\Big|_{(q_+p_-),(q_+l_-),y}.
\lab{amp-int}
\ea
Here $(q_+p_-), (q_+l_-), y$ are pole solutions of the $\delta$-functions,
the numerical factor has the form:
\be
cof = 4i\; \alpha_s^2 \; \alpha_e \; \frac{(-1)}{4 \pi^2} c_2,\;\;
 c_2 = \frac{8}{36}\;, \lab{cof}
\ee
$c_2$ is a color factor of the born two-gluon diagram (Fig. 1Á).
For the diagram investigated  the functions $G$ and $F$ look as follows:
\be
G(p\pm r) = - \frac{1-z}{m^2 + [p_\perp \pm r_\perp (1-z)]^2 },
\lab{prq}
\ee
\be
F(l \pm r) = - \frac{1}{\lambda^2 + [l_\perp \pm r_\perp ]^2 },\hspace{2cm}
 \lab{prg}
\ee
where we introduce the mass $\lambda$ in the  gluon propagators.
The matrix elements of the diagram have the following form:
\be
M_{Born}^{flip} = 8 s^2 \Delta^2 z^2 (z-1)^2
\ee
\be
M_{Born}^{non-flip} = \frac{4 s^2  z}{1-z} \left[\;
(2 z^4 - 6 z^3 + 7 z^2 - 4 z + 1)\Delta^2 - 4 m^2\right].
\ee
This means that the spin-flip and  spin-non-flip matrix elements
 are growing as $s^2$. One power of $s$ is compensated in the
integral (\ref{amp-int}) and both the spin-flip and non-flip amplitudes are
growing as  $s$; that is  taking the off-mass-shell effects into account
leads to the energy-independent ratio of spin-flip and spin-non-flip
amplitudes in the same $\alpha_s$ order.

The born two-gluon high-energy amplitude (Fig. 1a) has the form
\ci{gol5,go-se}:
\be
\hat T^{2g}(s,t)= A^{2g}(s,t) \frac{ \hat p'}{s},      \lab{t2g}
\ee
where
\be
 A^{2g}(s,t)=4is \alpha_s^2 c_2 \int d^2l_\perp \prod (F(l\pm r)),\;\;
 c_2 = \frac{8}{36},  \lab{a2g}
\ee
with $(F(l\pm r))$ determined in (\ref{prg}).\\
Note that the contribution of the radiative corrections in (\ref{t2g})
can be  taken into account through the use of the spinless pomeron
vertex function (formfactor)
\be
\frac{\mu _0\,^2}{\mu _0\,^2 + \Delta^2},\;\;\mu _0 \sim 1\,GeV,
\lab{ffk}
\ee
which was introduced in \ci{don-la}.
It is easy to see that  the integrals over $d^2p_\perp$
and $d^2l_\perp$ are factorized completely in (\ref{amp-int}). Moreover,
the integrals over
$d^2l_\perp$ coincide with the transverse integral in
(\ref{a2g}). So, expression (\ref{amp-int}) may be written in the form:
\ba
{\rm Im}\langle T_{Born}^i(s,t) \rangle=\alpha_e A^{2g}(s,t)\;\;
\Phi\,_{Born}^{\gamma\,q},    \lab{fact1}
\ea
where
\ba
\Phi\,_{Born}^{\gamma\,q}= \frac{(-1)}{4 \pi^2}
\int_{s_0/s}^{1}\frac{dz\;\left[M_{Born}^i \right]}{s^2\;z(1-z)}\;
\int\;d^2p_\perp\;\prod (G(p\pm r)).
\ea
 The obtained result (\ref{fact1}) confirm the
factorization in the spin-flip part of the pomeron exchange of the
large-distance effects $\Phi\,_{Born}^{\gamma\,q}$  and the
high energy spinless two gluon amplitude $A^{2g}(s,t)$.

The main contribution to the $\gamma q$-scattering amplitude in the
$\alpha_s^3$ order comes from the imaginary part of the diagrams drawn in
Figs. 2b, 2c.
We will also take into account the $\alpha_s^3$ order contributions to the
spin-non-flip amplitude in contrast with \ci{gol5,go-se}.
As it will be shown, the cross-section's shape depends essentially on these
contributions.
Let us calculate the imaginary parts of the spin-non-flip matrix elements
in the down quark line $\langle T_{Á,Â}^i(s,t) \rangle$ of the
diagrams Fig. 2b, 2c.  After integration with the $\delta$-functions
in the light-cone variables (\ref{lc}) we obtain the following
representation  similar to (\ref{amp-int}):
\ba
{\rm Im}\langle T_{1,2}^i(s,t) \rangle=cof_{1,2}\;\frac{1}{s}
\int_{s_0/s}^{1}\;\int_{s_0/s}^{z}\frac{dz\,dx}{(1-z)\,(z-x)\,x} \;\int\;
d^2p_\perp d^2k_\perp  d^2l_\perp \nonumber\\
\nonumber\\
\left[\;M_{1,2}^i
\;\prod (G(p\pm r)) \prod (G_{1,2}(k\pm r)) \prod(F(l\pm r))\right]\Big|_{
(q_+p_-),(q_+k_-),(q_+l_-),y}.
\lab{ampk-int}
\ea
Indices "1", "2"
denote that the quantity refers to graph Fig. 2b or 2c, respectively.
The numerical factors ($c_2$ was determined in (\ref{cof})) have the form:
\be
cof_1 = 4i\; \alpha_s^2 \; \alpha_e \; \frac{1}{(2 \pi)^4}\;3 c_2,\;\;
\ee
\be
cof_2 = 4i\; \alpha_s^2 \; \alpha_e \; \frac{1}{(2 \pi)^4}\;\frac{4}{3}c_2.
\ee
For the  diagrams investigated the functions $G$ and $F$ look as follows:
\be
G(p\pm r) = - \frac{1-z}{m^2 + [p_\perp \pm r_\perp (1-z)]^2 }\;,
\lab{propq}
\ee
\be
F(l \pm r) = - \frac{1}{\lambda^2 + [l_\perp \pm r_\perp ]^2 }\;,\hspace{2cm}
 \lab{propg}                                      \ee
$$
G_{1,2}(k\pm r) = -\frac{z-x}{z\left[a_{1,2} + (k_\perp \pm r_\perp b)^2
\right]}\;,
$$
\be
 b=\frac{z-x}{z}\,,\;\;
a_1=\frac{m^2x(1-x)}{z(1-z)}+ \frac{\lambda^2(z-x)}{z}+ \frac{x(z-x)(1-z)
r_\perp^2}{z^2}\;,
\lab{propq12}
\ee
$$
a_2=\frac{m^2x(z-x)(x-z+1)}{z(1-z)}+ \frac{\lambda^2x}{z}+ \frac{x(z-x)(1-z)
r_\perp^2}{z^2}\;.
$$
It follows from (\ref{propq}), (\ref{propg}), (\ref{propq12})
 that in the $\lambda \to 0$ limit we have
infrared singularities only from the
 gluon propagators $F(l\pm r)$ in the down part of the graph. All
other propagators do not have divergences in this limit.

The matrix elements of the diagrams have the following form:
\be
M_1^{flip} = 4 s^2 \Delta^2 m^2 x\;\Big[8x^2z-12x^2+2xz^2-8xz+5x+5z\Big]\;,
\ee
$$
M_1^{non-flip} = \frac{4 s^2 \Delta^2 m^2 x}{z(1-z)}
$$
\ba
\Big[\;8x^2z^3-20x^2z^2+16x^2z-4x^2+ 2xz^4\; -
\ea
$$
 10xz^3 + 19xz^2 -16xz +5z^3-10z^2+10z\Big]\;,
$$
\be
M_2^{flip} =- 16 s^2 \Delta^2 m^2 x^2\;\Big[2xz-3x-4z^2+8z-1\Big]\;,
\ee
\ba
M_2^{non-flip} = \frac{16 s^2 \Delta^2 m^2 x^2}{z(z-1)}
\ea
$$
 \Big[2xz^3-5xz^2+4xz-x-4z^4+12z^3-13z^2+6z\Big]\;.
$$
Hence, the spin-flip matrix elements and the spin-non-flip one of the
ladder diagrams are growing as $s^2$ as well. One power of $s$ is compensated
in the
integral (\ref{ampk-int}) and both the spin-flip and non-flip amplitudes are
growing as  $s$. The ratio of spin-flip and spin-non-flip amplitudes
is independent of the energy in the $\alpha_s^3$ order in the same manner as
for $\alpha_s^2$.
As one can see, the integrals over $d^2p_\perp$,  $d^2k_\perp$
and $d^2l_\perp$ are factorized completely in (\ref{ampk-int}).
Using (\ref{a2g}) as notation, the amplitudes (\ref{ampk-int}) can be written
in the form:
\ba
{\rm Im}\langle T_{1,2}^i(s,t) \rangle=\alpha_e A^{2g}(s,t)\;\;
\Phi\,_{1,2}^{\gamma\,q},    \lab{fact2}
\ea
where
\ba
\Phi\,_{1,2}^{\gamma\,q}= \frac{\alpha_s}{(2 \pi)^4}\;C_{1,2}
\int_{s_0/s}^{1}\;\int_{s_0/s}^{z}\frac{dz\,dx}{s^2\;(1-z)\,(z-x)\,x} \;
\left[\;M_{1,2}^i\;\right] \nonumber\\ \nonumber\\
\int\;d^2p_\perp \;\prod (G(p\pm r))\;\int\;d^2k_\perp\;\prod (G_{1,2}(k\pm r))
\ea
   $$  C_1=3,\;\;C_2=\frac{4}{3}. $$
 We have obtained again the factorization (\ref{fact2}) in the spin-flip part
of the pomeron exchange of the
large-distance effects $\Phi\,_{1,2}^{\gamma\,q}$ and the
high energy spinless two gluon amplitude $A^{2g}(s,t)$.

In calculation, we use $\alpha_s=0.3$ which is typical of
$|t|\sim 1GeV^2$ and $\lambda=0.1GeV$.
\\

The obtained results are shown in Fig. 3 and 4, where the $|t|$-dependence of
the amplitudes $\Phi\,^{\gamma\,q}$ is displayed. In Fig. 3, the contribution
of the born diagram subject to the formfactor (see (\ref{ffk})) is presented.
In Fig. 4, this contribution is shown without considering the formfactor.
It is easy to see that the cross-section's shape essentially depends on
the consideration of the formfactor and the
contributions in the $\alpha_s^3$ order to the spin-non-flip amplitude.
Actually, the total non-flip amplitude in Fig. 3 does not have
a dip in contrast with the
 one in Fig. 4. This fact means, respectively,
 the lack or the availability of diffractive minimum in the cross-section.
Hence, the obtained result may help to refine the formfactor expression
(\ref{ffk}) from the experimental data on the cross-section.

Let us consider the conclusions which may be drawn from the performed
calculation. As have been noted above, the spin-flip amplitudes are suppressed
logarithmically with respect to the spin-non-flip one for the quarks on the
mass shell and for the diagrams in the same $\alpha_s$ order.
As has been shown, the taking  off-mass-shell effects of the wave function
into account leads to the energy-independence of the ratio
$T_{flip}\Big/T_{non-flip}$ in $\alpha_s^2$ and $\alpha_s^3$ order.
One would expect that in higher $\alpha_s$ orders the energy-independence of
this function holds true. The next conclusion implies that  the wave function
 has no effect on the
factorization  of the large-distance contributions and the high energy
spinless two-gluon amplitude in $T_{flip}$
The most important conclusion to emerge from the obtained results is that
the spin-flip amplitude is not small with respect to the spin-non-flip one.
Its magnitude may run to 20-30\% of the spin-non-flip amplitude.
It should be emphasized that the obtained here spin-flip amplitudes are
completely determined by low-energy effects in the quark (gluon) loop.
Hence, the performed calculation shows that off-mass-shell effects
in the quark-loop essentially increase the contributions of the spin-dependent
quark-pomeron vertex to the amplitude with photon's spin-flip.
Such spin effects may lead to considerable spin asymmetries in the reactions
of the vector meson production which will be studied at future accelerators.
Note that it is necessary to find a relative  phase between spin-flip and
non-flip amplitudes
for the spin asymmetry computing.
These calculations and investigation of the diffractive photoproduction
of vector meson reactions will be performed later.

{\em Acknowledgement.}  The   authors   express   their deep  gratitude   to
 V.A.Meshcheryakov and A.N.Sissakian for the support
in this work and to A.V.Efremov, W.D.Nowak and O.V.Teryaev
for fruitful  discussions. O.L. would also like to thank V.G.Teplyakov.

    This work was supported in part by the Russian Fond of
Fundamental Research, Grant $   94-02-04616$.

\newpage
\begin{center}
{\bf Figure captions}
\end{center}
{\bf Fig.1} (a)- two-gluon $qq$-scattering diagram;
(b,c)- $\alpha_s^3$ contributions determining the spin-flip $qq$
-scattering amplitude. \\
{\bf Fig.2} The planar contributions to the $\gamma$q -scattering amplitude
(a)- born diagram; (b,c)- $\alpha_s^3$ contributions. \\
{\bf Fig.3} Different contributions to the  $\Phi^{\gamma\,q}$ spin-flip
amplitude -(a)
and spin-non-flip ampli-tude- (b): dot-dashed curves for the born (Fig. 2a)
diagram with the formfactor are taken into account;  dashed curves for the
ladder (Fig. 2b,c) diagrams; full line for the total
amplitude.\\
{\bf Fig.4} Different contributions to the  $\Phi^{\gamma\,q}$ spin-flip
amplitude -(a)
and spin-non-flip ampli-tude- (b): dot-dashed curves for the born (Fig. 2a)
diagram without the formfactor; dashed curves for the ladder
(Fig. 2b,c) diagrams; full line for the total amplitude.
\end{document}